\documentclass[preprint,superscriptaddress,amsmath,amssymb,aps,prd,longbibliography,floatfix,nofootinbib,11pt]{revtex4-1}
\usepackage{graphicx}   
\usepackage[
colorlinks=true,        
citecolor=blue,         
linkcolor=blue,         
urlcolor=blue          
]{hyperref}             
\usepackage{bm}         
\usepackage{xcolor}     
\usepackage{multirow}

\newcommand{\nc}{\newcommand*} 
\newcommand{\mU}{{\mathcal{U}}}

\nc{\Eq}[1]{Eq.~\eqref{#1}}     
\nc{\Fig}[1]{Fig.~\ref{#1}}     
\nc{\Table}[1]{Table~\ref{#1}}  
\nc{\Sec}[1]{Sec.~\ref{#1}}     
\def\({\left(}
\def\){\right)}
\def\[{\left[}
\def\]{\right]}
\def\e{\begin{equation}}
\def\q{\end{equation}}
\def\m{\begin{eqnarray}}
\def\n{\end{eqnarray}}

\begin{document}

\title{Detecting Cosmological Phase Transitions with Taiji: Sensitivity Analysis and Parameter Estimation}

\author{Fan Huang}
\affiliation{Institute of Theoretical Physics, Chinese Academy of Sciences, Beijing 100190, China}
\affiliation{School of Physical Sciences, University of Chinese Academy of Sciences, No. 19A Yuquan Road, Beijing 100049, China}

\author{Zu-Cheng Chen}
\email{zuchengchen@hunnu.edu.cn}
\affiliation{Department of Physics and Synergetic Innovation Center for Quantum Effects and Applications, Hunan Normal University, Changsha, Hunan 410081, China}
\affiliation{Institute of Interdisciplinary Studies, Hunan Normal University, Changsha, Hunan 410081, China}

\author{Qing-Guo Huang}
\email{huangqg@itp.ac.cn}
\affiliation{Institute of Theoretical Physics, Chinese Academy of Sciences, Beijing 100190, China}
\affiliation{School of Physical Sciences, University of Chinese Academy of Sciences, No. 19A Yuquan Road, Beijing 100049, China}
\affiliation{School of Fundamental Physics and Mathematical Sciences, Hangzhou Institute for Advanced Study, UCAS, Hangzhou 310024, China}


\begin{abstract}
We investigate the capability of the Taiji space-based gravitational wave observatory to detect stochastic gravitational wave backgrounds produced by first-order phase transitions in the early universe. Using a comprehensive simulation framework that incorporates realistic instrumental noise, galactic double white dwarf confusion noise, and extragalactic compact binary backgrounds, we systematically analyze Taiji's sensitivity across a range of signal parameters. Our Bayesian analysis demonstrates that Taiji can robustly detect and characterize phase transition signals with energy densities exceeding $\Omega_{\text{PT}} \gtrsim 1.4 \times 10^{-11}$ across most of its frequency band, with particularly strong sensitivity around $10^{-3}$ to $10^{-2}$ Hz. For signals with amplitudes above $\Omega_{\text{PT}} \gtrsim 1.1 \times 10^{-10}$, Taiji can determine the peak frequency with relative precision better than $10\%$. These detection capabilities would enable Taiji to probe electroweak-scale phase transitions in various beyond-Standard-Model scenarios, potentially revealing new physics connected to baryogenesis and dark matter production. We quantify detection confidence using both Bayes factors and the Deviance Information Criterion, finding consistent results that validate our statistical methodology.
\end{abstract}

\maketitle

\section{Introduction}

The direct detection of gravitational waves (GWs) by the LIGO and Virgo collaborations~\cite{LIGOScientific:2016aoc} has initiated a new era in observational astronomy, providing unprecedented access to astrophysical phenomena that remain invisible to electromagnetic observations. While ground-based detectors operate at frequencies between approximately 10 Hz and 1 kHz, space-based interferometers will explore the milli-Hertz frequency band, where signals from various cosmological sources are expected to be present~\cite{Caprini:2018mtu, Maggiore:2018sht}.

One of the potential targets for space-based GW observatories are stochastic GW backgrounds (SGWBs) produced by first-order phase transitions (FOPTs) in the early universe~\cite{Witten:1984rs,Hogan:1986dsh}. These transitions occur when a system transitions discontinuously between different vacuum states separated by an energy barrier, resulting in the nucleation and expansion of bubbles of the new phase within the old phase~\cite{Coleman:1977py,Linde:1981zj,Hindmarsh:2013xza, Hindmarsh:2015qta, Jinno:2016vai, Hindmarsh:2017gnf, Konstandin:2017sat, Cutting:2019zws, Pol:2019yex,Lewicki:2020jiv, Dahl:2021wyk,Jinno:2022mie,Auclair:2022jod,Sharma:2023mao,RoperPol:2023dzg}. In the Standard Model of particle physics, the electroweak phase transition is crossover-type; however, many well-motivated extensions predict a first-order electroweak phase transition occurring at temperatures of $T_* \sim 100$ GeV~\cite{Grojean:2006bp, Hindmarsh:2020hop}. Such phase transitions could explain the observed baryon asymmetry of the universe through electroweak baryogenesis~\cite{Kuzmin:1985mm, Cohen:1993nk}, and might be connected to dark matter production mechanisms~\cite{Baker:2019ndr}.

The Chinese space-based GW observatory Taiji~\cite{Hu:2017mde, Ruan:2018tsw} is one of several proposed missions designed to detect GWs in the milli-Hertz frequency range. Like the European Space Agency's Laser Interferometer Space Antenna (LISA)~\cite{LISA:2017pwj}, Taiji will consist of three spacecraft in a triangular formation, but with arm lengths of $3 \times 10^6$ km compared to LISA's $2.5 \times 10^6$ km. The Taiji constellation will follow a heliocentric orbit about $20^\circ$ ahead of Earth. Another Chinese space-based detector, TianQin~\cite{TianQin:2015yph}, is designed with shorter arm lengths of $\sim 10^5$ km and will be in Earth orbit, providing complementary sensitivity in partially overlapping frequency bands.

Detecting the SGWB from FOPTs requires distinguishing this cosmological signal from foreground sources, primarily from galactic and extragalactic compact binary (ECB) systems. The unresolved population of double white dwarf (DWD) binaries in our galaxy forms a significant confusion foreground~\cite{Farmer:2003pa, Ruiter:2007xx}, while the superposition of signals from ECB coalescences contributes an additional stochastic background~\cite{Zhu:2012xw, Rosado:2011kv}. The Taiji mission, with its specific noise characteristics and orbital configuration, presents unique capabilities and challenges for separating these components.

The detection of a SGWB from FOPTs faces significant challenges, primarily due to SGWBs contamination from unresolved galactic compact binaries, particularly DWD systems~\cite{Farmer:2003pa, Nelemans:2001hp}, and from extragalactic compact binaries~\cite{Regimbau:2011rp}. Those astrophysical SGWBs are strong enough that they become foregrounds acting as additional ``confusion noise" when conducting the detections of other GW signals in same frequency band~\cite{Romano:2016dpx}. Those foregrounds must be carefully modeled and subtracted to reveal primordial signatures~\cite{Cornish:2017vip,Kume:2024sbu}.

Previous studies have investigated LISA's capabilities to detect SGWBs from FOPTs~\cite{Caprini:2015zlo,Caprini:2019egz,Gowling:2021gcy,Gowling:2022pzb,Boileau:2022ter,Caprini:2024hue,Hindmarsh:2024ttn,Gonstal:2025qky}. More recently, attention has turned to the complementary capabilities of Taiji and the potential for joint observations with LISA or TinQin~\cite{Ruan:2019tje,Ruan:2020smc,Wang:2020dkc,Wang:2021srv,Wang:2023ltz,Jin:2023sfc,Cai:2023ywp,Liang:2024tgn}. However, a comprehensive analysis of Taiji's sensitivity to FOPTs, considering the latest noise models and foreground estimates, remains to be conducted.

In this paper, we comprehensively assess Taiji's capability to detect SGWBs from FOPTs. Our analysis incorporates detailed modeling of the Taiji noise spectrum, including both instrumental noise and astrophysical foreground contributions from galactic DWD binaries and ECB systems. We implement a Bayesian framework to systematically explore the detectability of phase transition signals across a range of amplitudes and peak frequencies, determining the regions of parameter space where Taiji can make robust detections and provide precise parameter estimates.
The paper is organized as follows: In Section~\ref{model}, we present our models for the GW signal from FOPTs, the Taiji detector sensitivity, and the relevant astrophysical foregrounds. Section~\ref{method} describes our Bayesian methodology and simulation framework. Finally, in Section~\ref{conclusion}, we summarize our findings and discuss their implications for probing beyond-Standard-Model physics with future space-based GW observatories.

\section{\label{model}Model components}
In this section, we describe the key components of our analysis framework. We first present our model for the GW signal from FOPTs, followed by a detailed characterization of the Taiji detector's noise properties. We then discuss the two primary astrophysical foregrounds that will impact the detection of cosmological signals: the galactic DWD confusion noise and the ECB background.

\subsection{SGWB from FOPTs}

For our analysis of FOPTs as sources of a SGWB, we adopt a simplified broken power-law spectral model obtained from fitting to numerical simulations~\cite{Hindmarsh:2017gnf}. The GW energy density is
\begin{equation} 
\Omega_{\text{GW}}(f) = \Omega_{\text{PT}}\, \mathcal{P}(f), 
\end{equation}
where the spectral shape function takes the form of
\begin{equation}\label{Pf}
\mathcal{P}(f) = \left(\frac{f}{f_{\text{PT}}}\right)^3 \left[\frac{7}{4 + 3(f/f_{\text{PT}})^2}\right]^{7/2}.
\end{equation}
Here, $\Omega_{\text{PT}}$ represents the peak amplitude of the SGWB, and $f_{\text{PT}}$ is the peak frequency~\cite{Hindmarsh:2017gnf}. The GW power spectrum relates to the power spectral density at the detector through
\begin{equation} 
\Omega_{\text{GW}}(f) = \frac{4\pi^2}{3H_0^{2}}f^3  S_{\text{PT}}(f), 
\end{equation}
where $H_0 = 67.4 \, \text{km} \, \text{s}^{-1} \, \text{Mpc}^{-1}$ is the Hubble parameter today~\cite{Planck:2018vyg}.
The peak frequency depends on the physical parameters of the phase transition:
\begin{equation} 
f_{\text{PT}} \simeq 10^{-6} (H_* R_*)^{-1} (T_*/100\text{ GeV})\text{ Hz},
\end{equation}
where $T_*$ is the temperature at which the phase transition occurs, $H_*$ is the Hubble rate at that time, and $R_*$ is the mean bubble separation.

The $f^3$ low-frequency behavior of $\mathcal{P}(f)$ in \Eq{Pf} is characteristic of phase transitions with mean bubble spacing on the order of the Hubble radius, which produce the strongest signals~\cite{Sharma:2023mao,RoperPol:2023dzg}. The high-frequency $f^{-4}$ behavior approximates the falloff seen in numerical simulations near the peak~\cite{Hindmarsh:2017gnf}. This model captures the essential features of FOPT signals while reducing the parameter space to two physically meaningful parameters: $\Omega_{\text{PT}}$ and $f_{\text{PT}}$.
For phase transitions in the temperature range of 100 GeV to 1 TeV (including the electroweak scale and many BSM scenarios), we expect peak frequencies between $10^{-4}$ Hz and $10^{-2}$ Hz with peak amplitudes in the range $10^{-14} < \Omega_{\text{PT}} < 10^{-9}$~\cite{Gowling:2021gcy}. These signals fall squarely within Taiji's sensitivity band, making Taiji a promising detector for probing BSM physics through GWs from FOPTs.

\subsection{Taiji noise model}

The Taiji space-based GW observatory features three spacecraft in a triangular configuration with 3 million kilometer arm lengths, longer than LISA's 2.5 million kilometers~\cite{Hu:2017mde,Ruan:2020smc}. To extract GW signals from the raw measurements, Taiji employs sophisticated signal processing techniques known as time delay interferometry (TDI)~\cite{Tinto:2001ii,Tinto:2002de}.
For our analysis, we focus on the interferometric data streams designated as the $X$, $Y$, and $Z$ TDI variables, which represent combinations of phase measurements that substantially reduce laser frequency noise. We adopt several simplifications in our noise modeling approach: 1) we assume that the SGWB signal and instrumental noise are uncorrelated, 2) we model the noise as consisting of two primary components, and 3) we treat all spacecraft as identical with equal arm lengths forming an equilateral triangle with $L = 3 \times 10^{6}$\,km~\cite{Luo:2020bls}.

The two dominant noise contributions in the Taiji detector can be characterized by their power spectral densities (PSDs). The first component arises from the optical measurement system (OMS), which dominates at higher frequencies (see \textit{e.g.}~\cite{Ren:2023yec})
\begin{equation} 
P_{\mathrm{oms}}(f) = P^2 \times 10^{-24} \frac{1}{\mathrm{Hz}}\left[1+\left(\frac{2\,\mathrm{mHz}}{f}\right)^4\right]\left(\frac{2 \pi f}{c} \mathrm{m}\right)^2, 
\end{equation}
where $P = 8$~\cite{Luo:2019zal}.
The second noise component comes from acceleration noise affecting the test masses, which dominates at lower frequencies
\begin{equation}
P_{\mathrm{acc}}(f) = A^2 \times 10^{-30} \frac{1}{\mathrm{Hz}}\left[1+\left(\frac{0.4\, \mathrm{mHz}}{f}\right)^2\right]\left[1+\left(\frac{f}{8\, \mathrm{mHz}}\right)^4\right]\left(\frac{1}{2 \pi f c} \frac{\mathrm{m}}{\mathrm{s}^2} \right)^2,
\end{equation}
where $A = 3$ characterizes the acceleration noise level~\cite{Luo:2019zal}.

With these noise components defined, we can express the noise auto-correlation in the $X$, $Y$, and $Z$ channels as
\begin{equation}\label{auto} 
N_{a a}(f, A, P)=16 \sin ^2\left(\frac{f}{f_*}\right)\left\{\left[3+\cos \left(\frac{2f}{f_*}\right)\right] P_{\mathrm{acc}}(f, A)+P_{\mathrm{oms}}(f, P)\right\}, 
\end{equation}
where $c$ is the speed of light and $f_*=c / (2 \pi L)$ defines a characteristic frequency of the detector geometry.
The cross-correlation between different channels (e.g., between $X$ and $Y$) is given by
\begin{equation}\label{cross}
N_{a b}(f, A, P)=-8 \sin ^2\left(\frac{f}{f_*}\right) \cos \left(\frac{f}{f_*}\right)\left[4 P_{\mathrm{acc}}(f, A)+P_{\mathrm{oms}}(f, P)\right].
\end{equation}

For analytical convenience, we transform the $X$, $Y$, and $Z$ channels into an alternative basis consisting of the channels A, E, and T
\begin{equation}
\left\{\begin{array}{l}
\mathrm{A}=\frac{1}{\sqrt{2}}(Z-X), \\
\mathrm{E}=\frac{1}{\sqrt{6}}(X-2 Y+Z), \\
\mathrm{T}=\frac{1}{\sqrt{3}}(X+Y+Z).
\end{array}\right.
\end{equation}
This transformation is advantageous because it produces noise-orthogonal channels $\mathrm{A}$ and $\mathrm{E}$ with identical noise properties, while $\mathrm{T}$ functions as a ``null channel" with reduced sensitivity to GWs~\cite{Prince:2002hp}. The noise power spectra in these channels can be derived as
\begin{equation}
N_{\mathrm{A,E}}=8 \sin ^2\left(\frac{f}{f_*}\right)  \left\{4\left[1+\cos \left(\frac{f}{f_*}\right)+\cos^2\left(\frac{f}{f_*}\right)\right] P_{\mathrm{acc}} +\left[2+\cos \left(\frac{f}{f_*}\right)\right] P_{\mathrm{oms}}\right\},
\end{equation}
and
\begin{equation}
N_{\mathrm{T}}=16 \sin ^2\left(\frac{f}{f_*}\right)  \left\{2\left[1-\cos \left(\frac{f}{f_*}\right)\right]^2 P_{\mathrm{acc}} + \left[1-\cos \left(\frac{f}{f_*}\right)\right] P_{\mathrm{oms}} \right\}. 
\end{equation}

To facilitate comparison with astrophysical and cosmological GW signals, we convert the noise spectral densities to equivalent energy spectral densities as
\begin{equation}\label{oa}
\Omega_{\alpha}(f)= S_{\alpha}(f) \frac{4 \pi^2 f^3}{3 H_0^2},
\end{equation}
where $\alpha \in \{\mathrm{A, E, T}\}$ denotes the channel, and $H_0$ is the Hubble constant. The noise spectral densities $S_\alpha(f)$ for each channel are defined as
\m
S_{\mathrm{A}}(f)&=&S_{\mathrm{E}}(f)=\frac{N_{\mathrm{A}}(f)}{\mathcal{R}_{\mathrm{A}}(f)},\\
S_{\mathrm{T}}(f)&=&\frac{N_{\mathrm{T}}(f)}{\mathcal{R}_{\mathrm{T}}(f)}.
\n
Here, $\mathcal{R}_\alpha$ corresponds to the response function for the respective channel $\alpha$. For this analysis, we employ the analytical expressions for these response functions as derived in~\cite{Wang:2021owg}.

\subsection{DWD foreground}

The Milky Way hosts a vast population of DWD binaries, with population synthesis models suggesting approximately $10^7-10^8$ such systems throughout our galaxy~\cite{Korol:2020lpq,Korol:2021pun}. These binaries generate gravitational radiation primarily in the frequency band spanning from $10^{-5}$ to $10^{-1}$ Hz~\cite{Karnesis:2021tsh}, which overlaps significantly with Taiji's detection window.

While Taiji will resolve individual signals from the strongest and closest sources, the vast majority of these binaries produce signals below the detection threshold. These unresolved systems generate a collective SGWB that manifests as an additional noise component in the detector, commonly referred to as the ``confusion noise" or ``galactic foreground"~\cite{Liu:2023qap}.
For a 4-year observation period, we approximate this galactic background using a broken power-law model
\begin{equation}
\Omega_{\mathrm{DWD}}(f)=\frac{A_1\left(f / f_*\right)^{\alpha_1}}{1+A_2\left(f / f_*\right)^{\alpha_2}}.
\end{equation}
The parameters that best fit the detailed population model are $A_1 = 3.98 \times 10^{-16}$, $A_2 = 4.79 \times 10^{-7}$, $\alpha_1 = -5.7$, and $\alpha_2 = -6.2$~\cite{Chen:2023zkb,Chen:2024jca}. This functional form captures the essential spectral features of the DWD background, particularly the high-frequency steepening that occurs as the number of contributing binaries decreases. This spectral break arises from physical constraints on binary orbital separations, which cannot be smaller than the combined radii of the component white dwarfs.
The corresponding energy density spectrum normalized to the critical density of the universe is given by
\begin{equation}
\Omega_{\mathrm{DWD}}(f)=S_{\mathrm{DWD}}(f) \frac{4 \pi^2 f^3}{3 H_0^2}.
\end{equation} 


\subsection{ECB foreground}

Beyond our galaxy, the universe contains innumerable compact binary systems that collectively generate a SGWB. This cosmological signal differs fundamentally from the galactic foreground, as it represents the superposition of unresolved binary black hole and neutron star systems distributed throughout cosmic history~\cite{Chen:2018rzo}.

While current ground-based interferometers have not yet reached the sensitivity required to detect this background, space-based detectors operating at lower frequencies will probe a different portion of its spectrum. The ECB background is particularly important for understanding the integrated merger history across cosmic time.

For our sensitivity analysis, we model this background with a characteristic power-law frequency dependence
\begin{equation}
\Omega_{\mathrm{ECB}}(f) = A_{\mathrm{ECB}}\left(\frac{f}{f_{\mathrm{ref}}}\right)^{\alpha_{\mathrm{ECB}}}.
\end{equation}
This spectral shape emerges naturally from the inspiral phase of compact binaries, with the $2/3$ power-law index reflecting the frequency evolution of binary systems dominated by gravitational radiation. We adopt an amplitude of $A_{\mathrm{ECB}} = 1.8 \times 10^{-9}$ at the reference frequency $f_{\mathrm{ref}} = 25$ Hz~\cite{Chen:2018rzo}.

Unlike the galactic foreground, this background exhibits no spectral breaks within the Taiji frequency band, as the contributing sources span a much broader range of masses, redshifts, and formation channels.

\section{\label{method}Methodology and Results}


\begin{figure}[htbp!]
	\centering
	\includegraphics[width=0.8\linewidth]{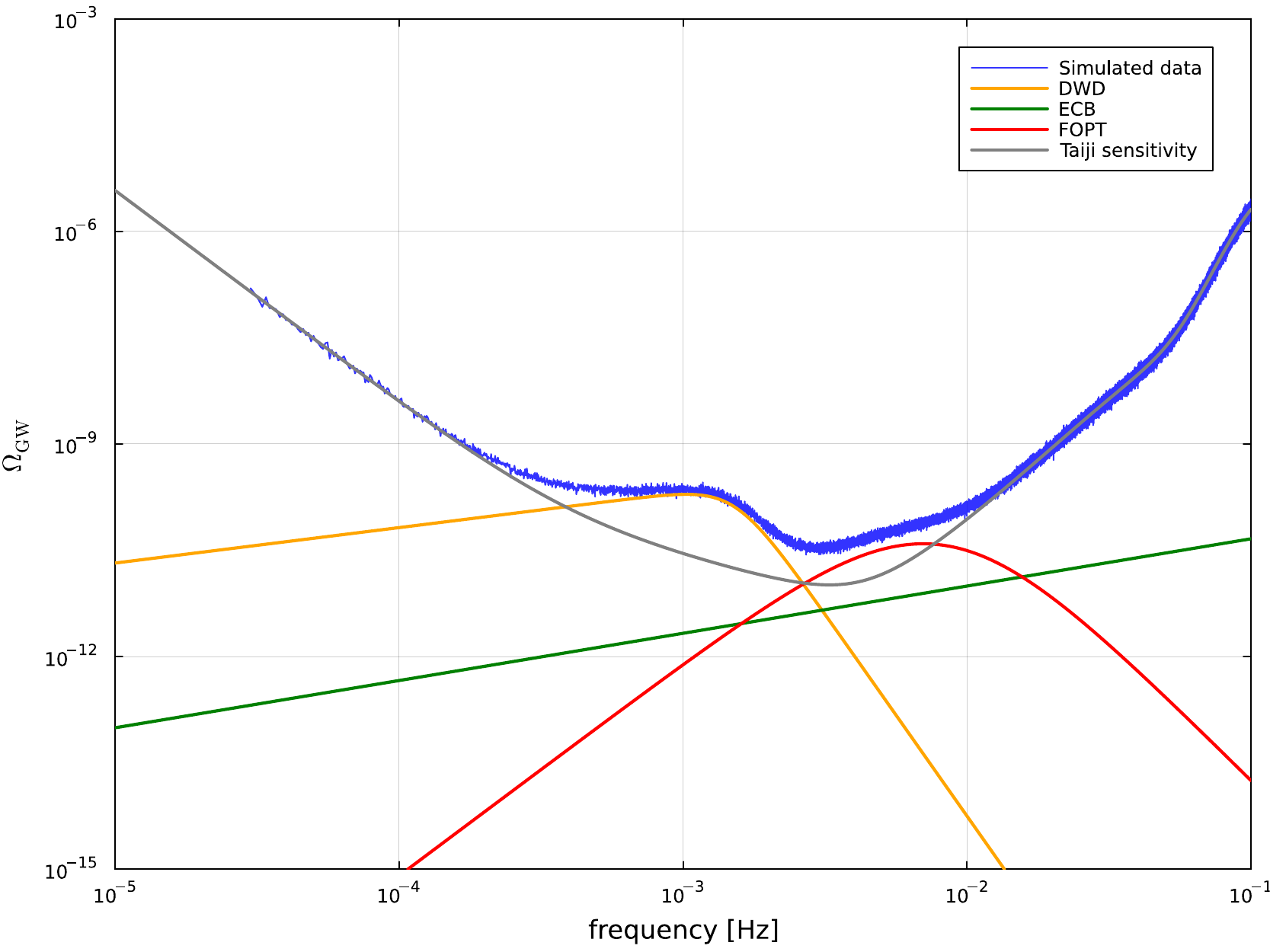}
	\caption{\label{data}Frequency-domain representation of synthetic Taiji A-channel observations (blue). We also show the galactic DWD confusion noise (orange), the contribution from ECB (green), and the cosmological background from the FOPT (red) with parameters $\Omega_{\mathrm{PT}}=3.9\times 10^{-11}$ and peak frequency $f_{\mathrm{PT}}= 7\times 10^{-3} \mathrm{Hz}$. For reference, the Taiji detector's sensitivity is plotted in terms of $\Omega_{\mathrm{GW}}(f)$ as a gray curve. 
 }
\end{figure}

This section outlines our computational approach for evaluating the Taiji mission's capability to detect SGWBs from cosmological FOPTs following~\cite{Caprini:2019pxz,Flauger:2020qyi}.  
Our numerical framework simulates Taiji observations spanning the full 4-year mission duration with realistic duty cycle considerations (assuming 75\% efficiency~\cite{Caprini:2019pxz,Seoane:2021kkk,Wang:2021njt}), yielding an effective 3-year observations. We segment the TDI measurements into roughly $N_c = 94$ chunks of 11.5 days each~\cite{Caprini:2019pxz,Flauger:2020qyi}. The frequency domain extends from $3 \times 10^{-5}$ Hz to $0.5$ Hz with approximately $5 \times 10^7$ total data points at $10^{-6}$ Hz resolution.

For computational implementation, we transform the time-domain signal into frequency space:
\begin{equation}
d(t) = \sum_{f=f_{\min}}^{f_{\max}} \left[d(f) e^{-2 \pi i f t} + d^*(f) e^{2 \pi i f t}\right].
\end{equation}
Under the assumption of stationarity for both signal and noise components, the Fourier coefficients exhibit the following statistical properties:
\begin{equation}
\left\langle d(f) d\left(f^{\prime}\right)\right\rangle=0 \quad \text { and } \quad\left\langle d(f) d^*\left(f^{\prime}\right)\right\rangle=D(f) \delta_{f f^{\prime}},
\end{equation}

\begin{table}
\centering
\begin{tabular}{c|c|c|c}
\hline\hline
Parameter & Prior & Injected value  & Recovered value \\ 
\hline
$A$ & $\mU(2.95, 3.05)$ & $3$  & $3.002^{+0.007}_{-0.007}$ \\ 
$P$ & $\mU(7.99, 8.01)$ & $8$  & $7.9990^{+0.0013}_{-0.0014}$ \\ 
$\log_{10} A_1$ & $\mU(-16, -15)$ & $-15.4$  & $-15.39^{+0.04}_{-0.05}$ \\ 
$\alpha_1$ & $\mU(-6, -5.5)$ & $-5.7$  & $-5.69^{+0.05}_{-0.05}$ \\ 
$\log_{10} A_2$ & $\mU(-6.5, -6)$  & $-6.32$  & $-6.31^{+0.04}_{-0.04}$ \\ 
$\alpha_2$ & $\mU(-6.5, -6)$ & $-6.2$  &$-6.19^{+0.04}_{-0.04}$ \\ 
$\log_{10} A_{\mathrm{ECB}}$ & $\mU(-9, -8.5)$ & $-8.74$  & $-8.69^{+0.15}_{-0.18}$ \\ 
$\alpha_{\mathrm{ECB}}$ & $\mU(0.5, 1)$ & $2/3$  & $0.68^{+0.04}_{-0.05}$ \\ 
$\log_{10} \Omega_{\mathrm{PT}}$ & $\mU(-10.609, -10.209)$ & $-10.409$  & $-10.408^{+0.004}_{-0.004}$ \\ 
$\log_{10} (f_{\mathrm{PT}}/\mathrm{Hz})$ & $\mU(-2.355, -1.955)$ & $-2.155$  & $-2.154^{+0.002}_{-0.001}$ \\ 
\hline
\end{tabular}
\caption{\label{tab:priors}Summary of Bayesian analysis results for all model parameters. The table displays the uniform prior ranges ($\mU$) employed in our MCMC sampling, alongside the true parameter values used in synthetic data generation. The rightmost column presents the posterior estimates, showing median values with corresponding 90\% credible intervals. 
The prior ranges are chosen to balance computational efficiency with statistical robustness while focusing on the theoretically motivated parameter space for FOPTs detectable by Taiji.
}
\end{table}

The simulation generates synthetic observations by drawing complex Fourier coefficients from Gaussian distributions characterized by the appropriate power spectral densities. Specifically, at each frequency point, we construct: 
\m
S_i & =&\left|\frac{G_{i 1}\left(0, \sqrt{\Omega_{\mathrm{GW}}\left(f_i\right)}\right)+i G_{i 2}\left(0, \sqrt{\Omega_{\mathrm{GW}}\left(f_i\right)}\right)}{\sqrt{2}}\right|^2, \\
N_i & =&\left|\frac{G_{i 3}\left(0, \sqrt{\Omega_{\mathrm{A,E,T}}\left(f_i\right)}\right)+i G_{i 4}\left(0, \sqrt{\Omega_{\mathrm{A,E,T}}\left(f_i\right)}\right)}{\sqrt{2}}\right|^2 .
\n
Here, $G_{ij}(M,\sigma)$ represents random samples from a Gaussian distribution with mean $M$ and standard deviation $\sigma$. The total power at each frequency combines signal and noise contributions: $D_i = S_i + N_i$. To account for statistical fluctuations, we generate $N_\mathrm{c}$ independent realizations $\{D_{i1}, D_{i2}, \ldots, D_{iN_\mathrm{c}}\}$ at each frequency and compute their ensemble average $\bar{D}_i$. \Fig{data} illustrates a representative simulated dataset, with injection parameters documented in Table \ref{tab:priors}.

\begin{figure}[tbp!]
	\centering
	\includegraphics[width=\linewidth]{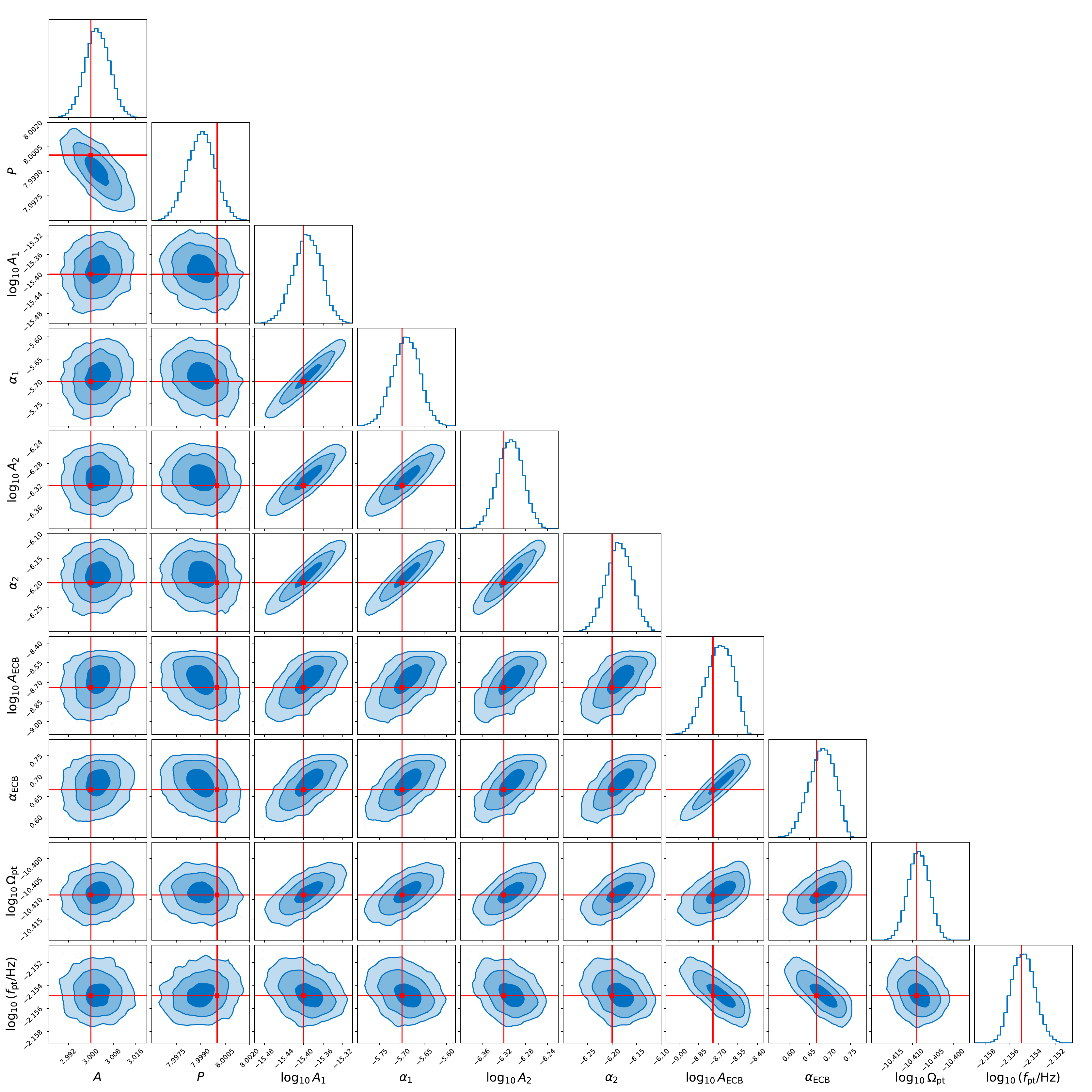}
	\caption{\label{posts_Taiji}    
    Posterior distributions of model parameters from Bayesian analysis using simulated Taiji data. The corner plot shows marginalized one-dimensional posteriors along the diagonal and joint two-dimensional distributions with $1\sigma$, $2\sigma$, and $3\sigma$ confidence contours in the off-diagonal panels. Red markers indicate the true parameter values used in generating the synthetic signal, which featured a FOPT with amplitude $\Omega_{\mathrm{PT}}=3.9\times 10^{-11}$ and characteristic frequency $f_{\mathrm{PT}}= 7\times 10^{-3} \mathrm{Hz}$.}
\end{figure}

To enhance computational efficiency while preserving information content, we implement adaptive frequency binning. For frequencies below $10^{-3}$ Hz, we maintain the original resolution, while frequencies between $10^{-3}$ Hz and $0.5$ Hz are rebinned into 1000 logarithmically spaced intervals. This optimization reduces the dataset to 1971 frequency bins per segment. The rebinned data is calculated as: 
\m
f_{(k)} &\equiv& \sum_{j \in \operatorname{bin} k} w_j f_j, \\
D_{(k)} &\equiv& \sum_{j \in \operatorname{bin} k} w_j \bar{D}_j,
\n
where the optimal weights are 
\begin{equation}\label{wj}
w_j =\frac{\mathcal{D}^\mathrm{th}(f_j, \vec{\theta}, \vec{n})^{-1}}{\sum_{l \in \operatorname{bin} k}\mathcal{D}^\mathrm{th}(f_l, \vec{\theta}, \vec{n})^{-1}}.
\end{equation}
Here, $\mathcal{D}^\mathrm{th}(f_j, \vec{\theta}, \vec{n}) \equiv \Omega_{\mathrm{GW}} (\vec{\theta}, f_j) + \Omega_{\alpha} (\vec{n}, f_i)$ represents the theoretical model for the total energy density, which is an estimate of the variance of the segment-averaged data $\bar{D}_j$~\cite{Flauger:2020qyi}. The parameter $\vec{n} \equiv \{\mathcal{A}, P\}$ denotes the instrumental noise parameters, while $\vec{\theta} \equiv \{A_1, \alpha_1, A_2, \alpha_2, A_\mathrm{ECB}, \alpha_\mathrm{ECB}, \Omega_\mathrm{PT}, f_\mathrm{PT}\}$ encompasses all astrophysical and cosmological signal parameters, including the galactic DWD foreground, ECB background, and the FOPT signal of interest. 

We now provide a brief derivation of \Eq{wj}. Each data point $\bar{D}_j$ has variance $\text{Var}(\bar{D}_j) = D_{\text{th}}(f_j, \vec{\theta}, \vec{n})$.
For the binned estimator $D_{(k)} = \sum_{j \in \text{bin}\,k} w_j \bar{D}_j$, assuming uncorrelated data points within each bin, the variance is 
\begin{equation}
    \text{Var}(D_{(k)}) = \sum_{j \in \text{bin}\,k} w_j^2 \, \text{Var}(\bar{D}_j) = \sum_{j \in \text{bin}\,k} w_j^2 \, D_{\text{th}}(f_j, \vec{\theta}, \vec{n}).
\end{equation}
To minimize this variance subject to the normalization constraint 
\begin{equation}\label{norm}
    \sum_{j \in \text{bin}\, k} w_j = 1,
\end{equation}
we use the method of Lagrange multipliers.
The Lagrangian is 
\begin{equation}
    \mathcal{L} = \sum_{j \in \text{bin}\, k} w_j^2 \, D_{\text{th}}(f_j, \vec{\theta}, \vec{n}) - \lambda \left(\sum_{j \in \text{bin}\,k} w_j - 1\right).
\end{equation}
Taking the derivative with respect to $w_i$ and setting it to zero yields
\begin{equation}\label{wi2}
    w_i = \frac{\lambda}{2 D_{\text{th}}(f_i, \vec{\theta}, \vec{n})}.
\end{equation}
Applying the normalization constraint in \Eq{norm}, we obtain 
\begin{equation}\label{lambda}
    \lambda = \frac{2}{\sum_{j \in \text{bin}\,k} D_{\text{th}}(f_j, \vec{\theta}, \vec{n})^{-1}}.
\end{equation}
Substituting \Eq{lambda} back yields the optimal weights in \Eq{wj}.

Our statistical analysis employs a hybrid likelihood function combining Gaussian and log-normal components~\cite{Flauger:2020qyi}, namely
\begin{equation}
\ln \mathcal{L}=\frac{1}{3} \ln \mathcal{L}_\mathrm{G}+\frac{2}{3} \ln \mathcal{L}_\mathrm{LN}.
\end{equation}
The Gaussian part is
\begin{equation}
\ln \mathcal{L}_\mathrm{G}(D \mid \vec{\theta}, \vec{n})=-\frac{N_{\mathrm{c}}}{2} \sum_{\alpha} \sum_k n_{\alpha}^{(k)}\left[\frac{\mathcal{D}_{\alpha}^\mathrm{t h}\left(f_{\alpha}^{(k)}, \vec{\theta}, \vec{n}\right)-\mathcal{D}_{\alpha}^{(k)}}{\mathcal{D}_{\alpha}^\mathrm{t h}\left(f_{\alpha}^{(k)}, \vec{\theta}, \vec{n}\right)}\right]^2,
\end{equation}
while the log-normal part is 
\begin{equation}
\ln \mathcal{L}_\mathrm{L N}(D \mid \vec{\theta}, \vec{n})=-\frac{N_{\mathrm{c}}}{2} \sum_{\alpha} \sum_k n_{\alpha}^{(k)} \ln ^2\left[\frac{\mathcal{D}_{\alpha}^\mathrm{t h}\left(f_{\alpha}^{(k)}, \vec{\theta}, \vec{n}\right)}{\mathcal{D}_{\alpha}^{(k)}}\right].
\end{equation}
The inclusion of the log-normal component in our likelihood function is crucial for properly handling the statistical properties of power spectral densities. When analyzing SGWB signals, the power spectral densities follow a $\chi^2$ distribution rather than a Gaussian distribution. Using solely a Gaussian likelihood in such cases can lead to biased parameter estimation, particularly for weak signals where the signal-to-noise ratio is low~\cite{Flauger:2020qyi}. The log-normal term better captures the right-skewed nature of the $\chi^2$ distribution while maintaining computational tractability. This hybrid likelihood approach has been widely adopted and validated in the literature for SGWB analyses (see e.g.~\cite{Flauger:2020qyi,Dimitriou:2023knw,Kume:2024sbu}).

To quantitatively assess the detectability of phase transition signals, we employ two complementary model selection metrics: the Bayes factor (BF) and the Deviance Information Criterion (DIC). The Bayes factor represents the ratio of evidences between competing models, providing a direct measure of relative model probability. Specifically, we define BF as
\begin{equation}
    \mathrm{BF} = \frac{\mathcal{Z}_\text{FOPT}}{\mathcal{Z}_\text{null}},
\end{equation}
where $\mathcal{Z}_\text{FOPT}$ is the evidence for the model including a PT component and $\mathcal{Z}_\text{null}$ represents the model with only astrophysical foregrounds and instrumental noise. Values of $\ln (\mathrm{BF}) > 8$ indicate decisive evidence favoring the presence of a phase transition signal. As a complementary approach, the DIC incorporates both goodness-of-fit and model complexity through 
\begin{equation}
    \text{DIC} = D(\bar{\theta}) + 2 p_D,
\end{equation}
where $\bar{\theta}$ represents the posterior mean, $D(\theta) = -2\ln\mathcal{L}(\theta)$, and $p_D = \bar{D(\theta)} - D(\bar{\theta})$ is the penalization term.
The difference $\Delta\text{DIC} = \text{DIC}_\text{null} - \text{DIC}_\text{FOPT}$ provides another measure of model preference, with larger positive values supporting the inclusion of the phase transition component.

Parameter estimation is performed using the nested sampling algorithm implemented in \texttt{dynesty}, accessed through the \texttt{Bilby} Bayesian inference library. Figure \ref{posts_Taiji} displays the resulting posterior distributions for a representative FOPT signal with amplitude $\Omega_{\mathrm{PT}}=3.9\times 10^{-11}$ and characteristic frequency $f_{\mathrm{PT}}= 7\times 10^{-3} \mathrm{Hz}$. The recovered values, along with their median and $90\%$ equal-tail uncertainties, are also summarized in \Table{tab:priors}.

\begin{figure}[htbp!]
	\centering
	\includegraphics[width=0.9\linewidth]{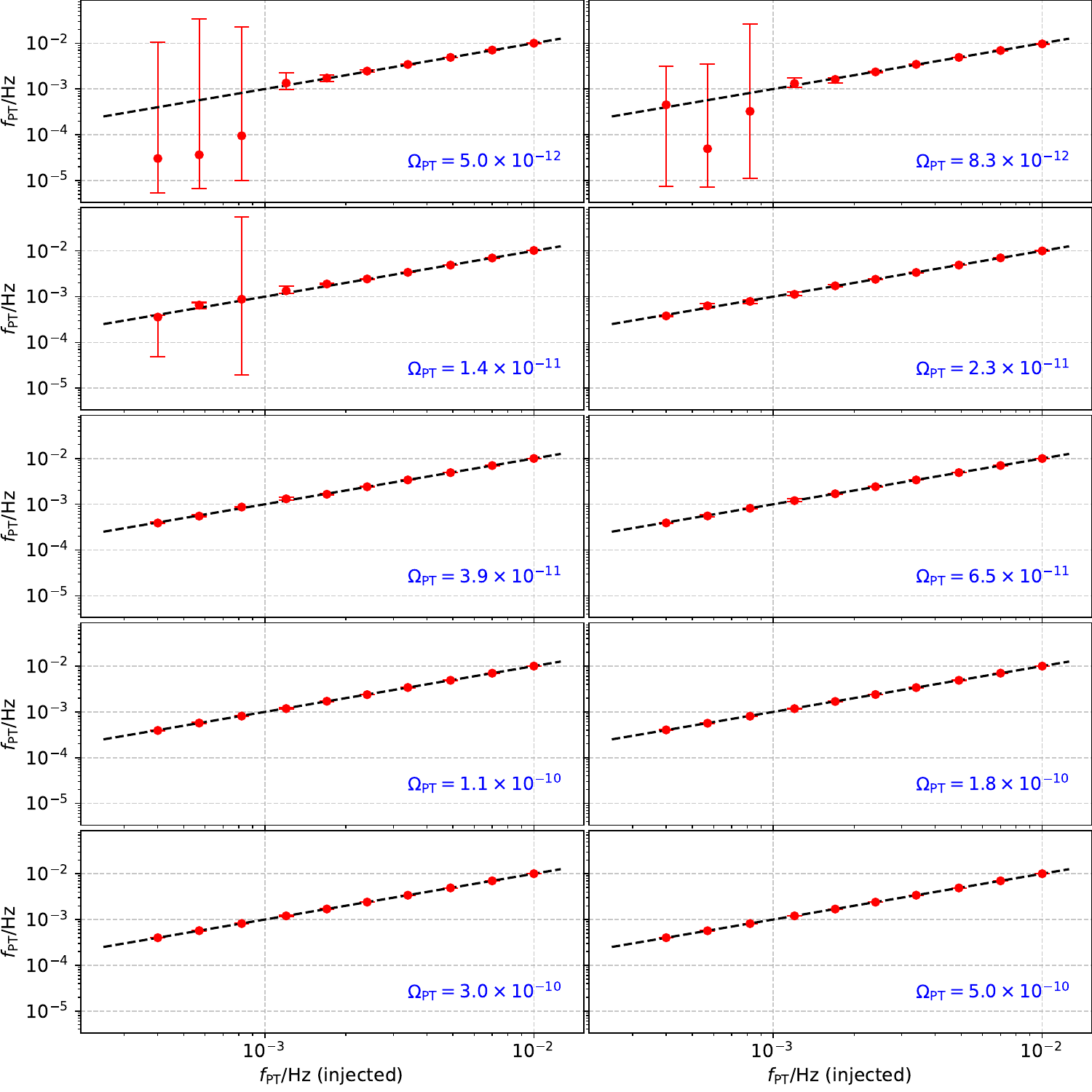}
	\caption{\label{fpts}Comparison between injected and recovered peak frequencies ($f_\mathrm{PT}$) of the FOPT signal. Each point represents the median of the posterior distribution, with error bars indicating the 90\% credible intervals. The dashed line represents perfect recovery.}
\end{figure}

\begin{figure}[htbp!]
	\centering
	\includegraphics[width=0.9\linewidth]{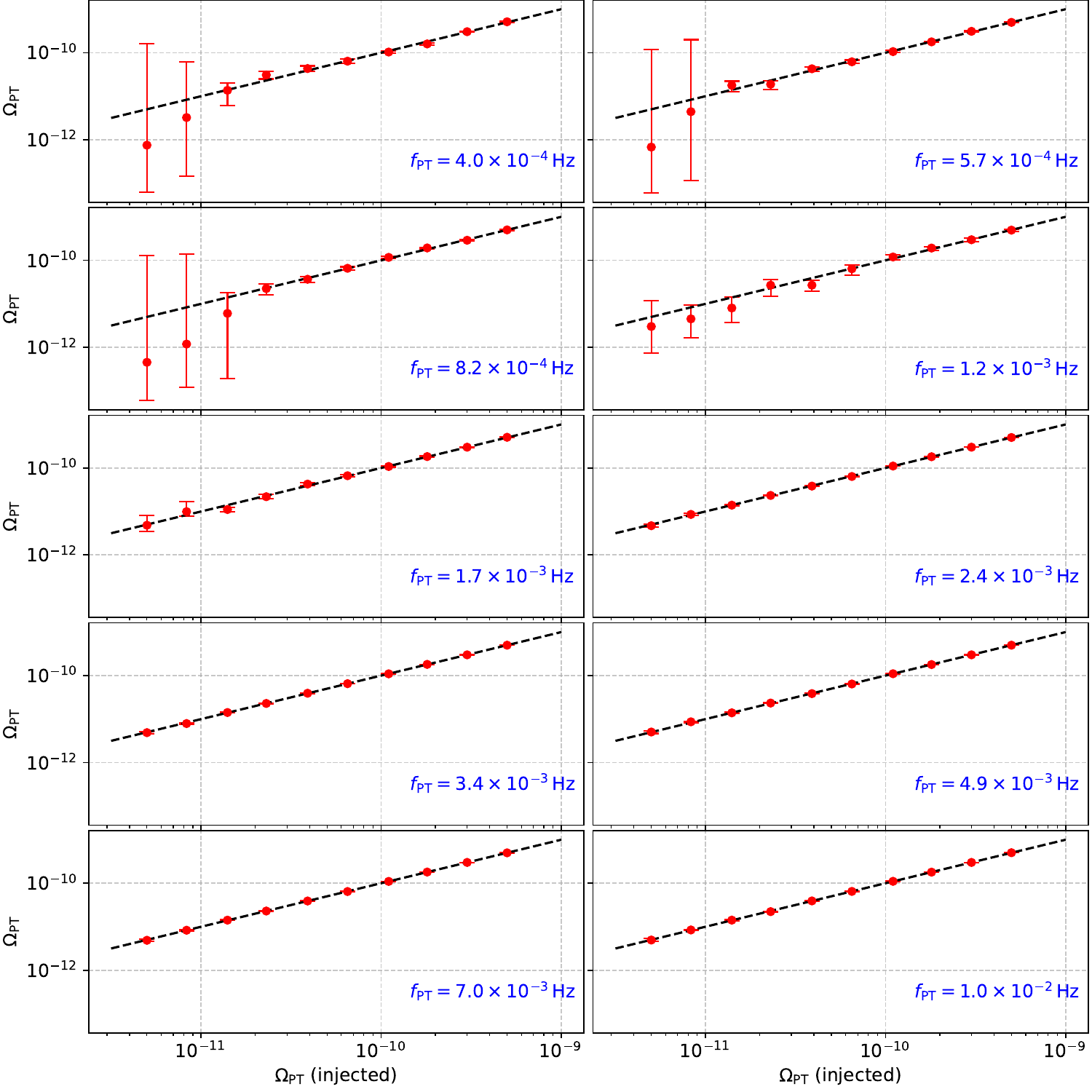}
	\caption{\label{Opts}Comparison between injected and recovered amplitudes ($\Omega_\mathrm{PT}$) of the FOPT signal. Each point represents the median of the posterior distribution, with error bars indicating the 90\% credible intervals. The dashed line represents perfect recovery.}
\end{figure}

\begin{figure}[htbp!]
	\centering
	\includegraphics[width=0.8\linewidth]{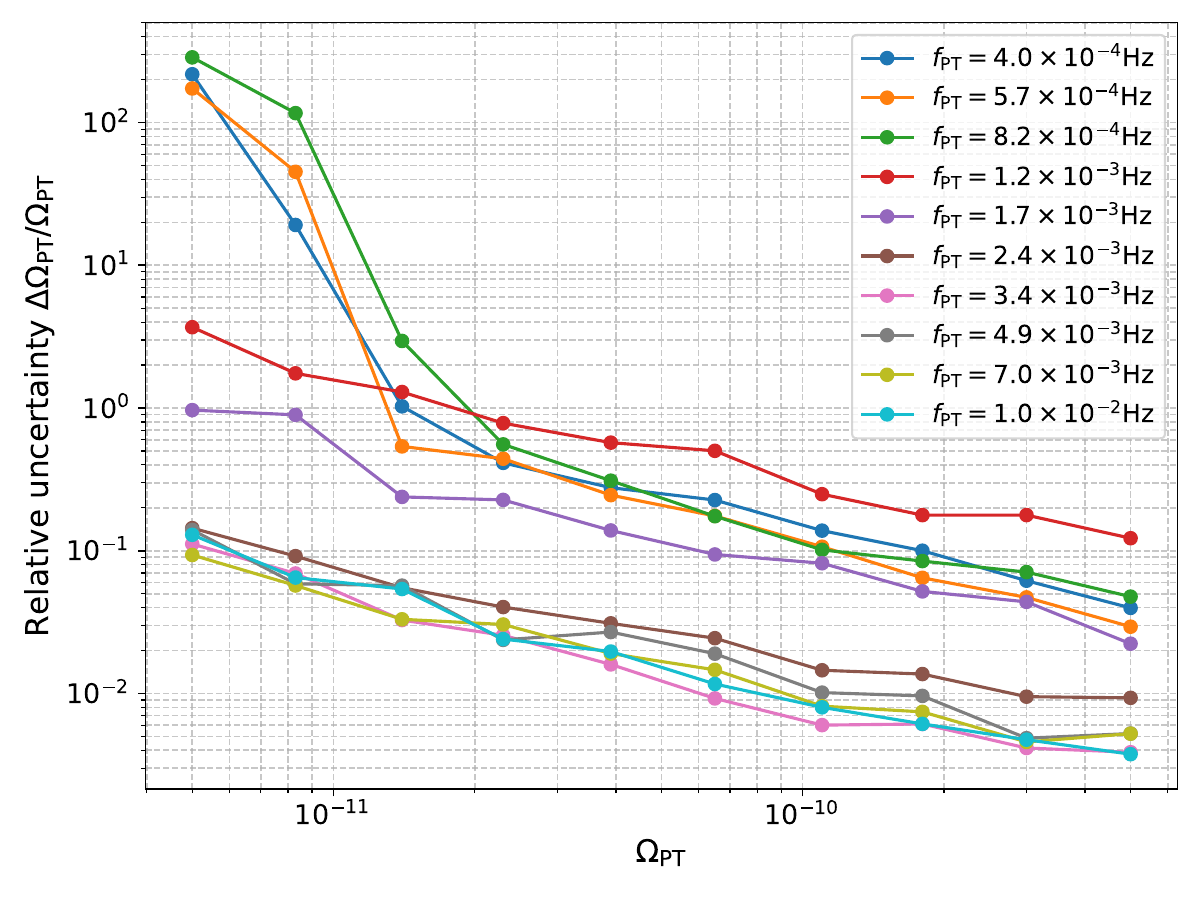}
	\caption{\label{dOpt}Measurement precision of the phase transition amplitude as a function of signal strength. The vertical axis shows the relative uncertainty ($\Delta\Omega_\mathrm{PT}/\Omega_\mathrm{PT}$) in the recovered amplitude, while the horizontal axis displays the injected amplitude values.}
\end{figure}

\begin{figure}[htbp!]
	\centering
	\includegraphics[width=0.8\linewidth]{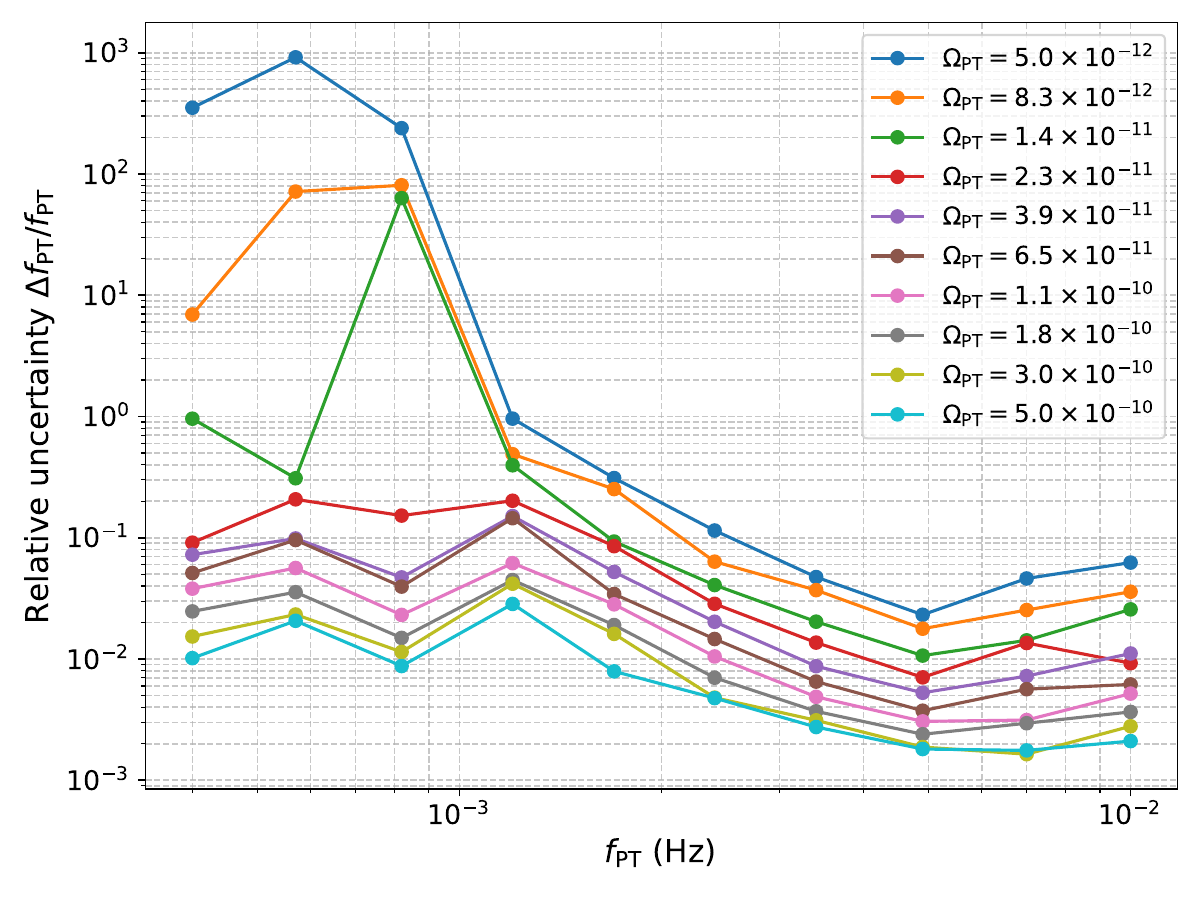}
	\caption{\label{dfpt}Frequency resolution capabilities of the analysis pipeline across the detection band. The plot shows the relative uncertainty ($\Delta f_\mathrm{PT}/f_\mathrm{PT}$) in peak frequency estimation as a function of the injected signal frequency.}
\end{figure}

\begin{figure}[htbp!]
	\centering
	\includegraphics[width=0.8\linewidth]{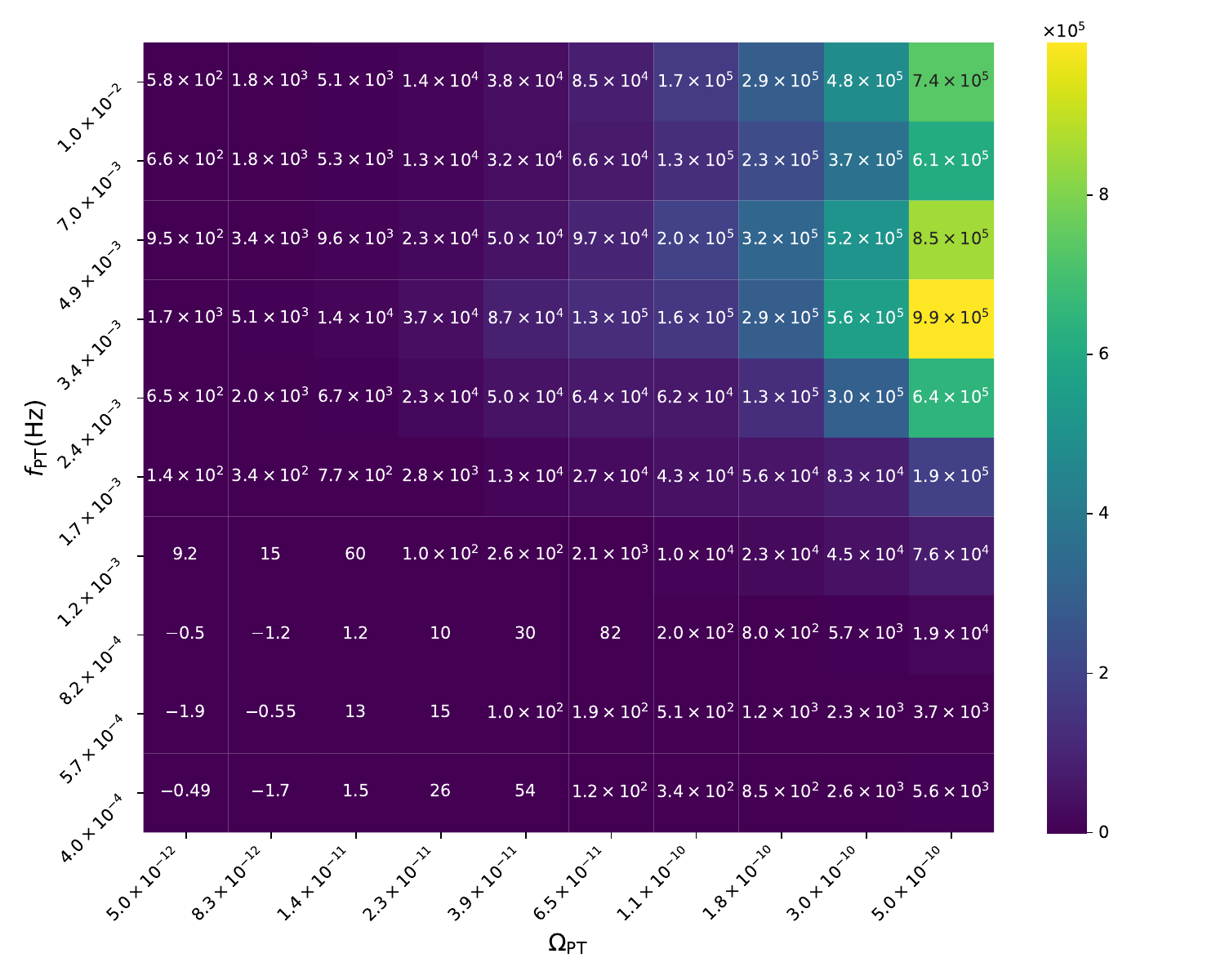}
	\caption{\label{BF_heatmap}Model selection analysis using the Bayes factors. The heatmap displays logarithmic Bayes factors comparing models with and without a phase transition component, plotted as a function of signal amplitude ($\Omega_\mathrm{PT}$) and peak frequency ($f_\mathrm{PT}$).}
\end{figure}

\begin{figure}[htbp!]
	\centering
	\includegraphics[width=0.8\linewidth]{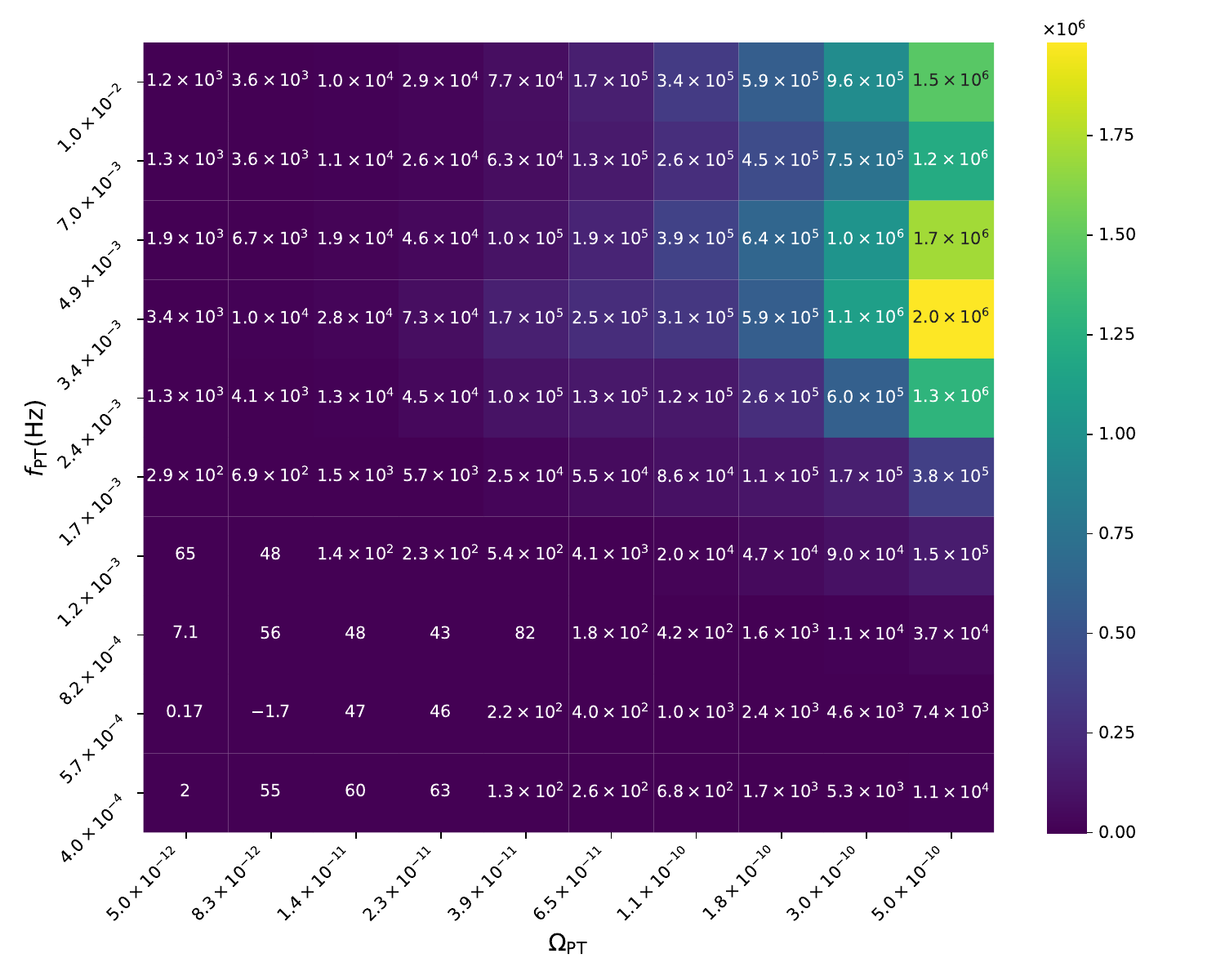}
	\caption{\label{DIC_heatmap}Model selection analysis using the Deviance Information Criterion (DIC). The heatmap illustrates the difference in DIC values between models with and without a phase transition component across the parameter space of signal amplitude ($\Omega_\mathrm{PT}$) and characteristic frequency ($f_\mathrm{PT}$).}
\end{figure}

Our simulation framework incorporates a set of base parameters, including: the detector noise characterization parameters fixed at reference values of $A=3$ and $P=8$; Galactic foreground modeling with four parameters describing the DWD confusion noise: amplitude coefficients $A_1 = 3.98 \times 10^{-16}$ and $A_2 = 4.79 \times 10^{-7}$, with corresponding spectral slopes $\alpha_1 = -5.7$ and $\alpha_2 = -6.2$; ECB background parameterized by amplitude $A_{\mathrm{ECB}} = 1.8 \times 10^{-9}$ with canonical spectral index $\alpha_{\mathrm{ECB}}=2/3$. While these parameters remain constant throughout our analysis, it is important to note that each simulation represents a distinct statistical realization of the stochastic backgrounds, as the foreground components are characterized by their power spectral densities rather than deterministic waveforms. 

Against this realistic background, we systematically inject phase transition signals spanning a two-dimensional parameter grid. The signal strength parameter $\Omega_{\mathrm{PT}}$ and characteristic frequency $f_{\mathrm{PT}}$ are varied across the following ranges:
\begin{equation*}
\begin{aligned}
\Omega_{\mathrm{PT}} \in & \left\{5.0 \times 10^{-12}, 8.3 \times 10^{-12}, 1.4 \times 10^{-11}, 2.3 \times 10^{-11}, 3.9 \times 10^{-11}, 6.5 \times 10^{-11},\right. \\
& \left. 1.1 \times 10^{-10}, 1.8 \times 10^{-10}, 3.0 \times 10^{-10}, 5.0 \times 10^{-10}\right\}, \\
f_{\mathrm{PT}} / \mathrm{Hz} \in & \left\{4.0 \times 10^{-4}, 5.7 \times 10^{-4}, 8.2 \times 10^{-4}, 1.2 \times 10^{-3}, 1.7 \times 10^{-3}, 2.4 \times 10^{-3},\right. \\
& \left. 3.4 \times 10^{-3}, 4.9 \times 10^{-3}, 7.0 \times 10^{-3}, 1.0 \times 10^{-2}\right\}.
\end{aligned}
\end{equation*}
This parameterization creates a grid of 100 distinct signal configurations, each requiring a separate Markov Chain Monte Carlo (MCMC) analysis. \Fig{data} illustrates the frequency-domain representation of synthetic Taiji data for a representative case with $\Omega_{\mathrm{PT}}=3.9\times 10^{-11}$ and $f_{\mathrm{PT}}= 7\times 10^{-3} \mathrm{Hz}$. The corresponding posterior distributions for this benchmark scenario are presented in \Fig{posts_Taiji}, demonstrating that all model parameters are successfully recovered within the $2\sigma$ credible intervals.

\Fig{fpts} and \Fig{Opts} display the measurement uncertainties in the recovered peak frequency $f_{\mathrm{PT}}$ and amplitude $\Omega_{\mathrm{PT}}$, respectively, across the parameter space. The error bars exhibit significant growth when $\Omega_{\mathrm{PT}}$ falls below $1.4\times 10^{-11}$ or when $f_{\mathrm{PT}}$ is less than $1.2 \times 10^{-3} \mathrm{Hz}$. This degradation in parameter estimation precision can be attributed to the competing influence of the DWD confusion background, which dominates the detector's low-frequency sensitivity band and effectively masks cosmological signals below certain amplitude thresholds in this frequency regime.

\Fig{dOpt} presents the relative uncertainty in amplitude ($\Delta\Omega_\mathrm{PT}/\Omega_\mathrm{PT}$) while \Fig{dfpt} illustrates the relative uncertainty in peak frequency ($\Delta f_\mathrm{PT}/f_\mathrm{PT}$) across the parameter space. As expected, $\Delta\Omega_\mathrm{PT}/\Omega_\mathrm{PT}$ demonstrates a clear inverse relationship with signal strength, decreasing systematically as $\Omega_\mathrm{PT}$ increases due to improved signal-to-noise ratio. Similarly, the fractional uncertainty in frequency determination $\Delta f_\mathrm{PT}/f_\mathrm{PT}$ also diminishes with increasing signal amplitude. Notably, when the phase transition signal reaches $\Omega_\mathrm{PT} \gtrsim 1.1\times 10^{-10}$, the frequency can be determined with high precision, achieving $\Delta f_\mathrm{PT}/f_\mathrm{PT} \lesssim 0.1$ across most of the frequency range.

To quantitatively evaluate model selection capabilities, we present the logarithmic BFs in \Fig{BF_heatmap} and the DIC differences in \Fig{DIC_heatmap}, comparing models with and without the phase transition component across the parameter space. Both metrics exhibit consistent behavior, showing progressive improvement in detection confidence as $\Omega_\mathrm{PT}$ increases. This concordance between independent statistical measures reinforces our confidence in the results. The observed trend aligns with theoretical expectations, as larger amplitude signals naturally produce more decisive evidence for the presence of a cosmological phase transition against the null hypothesis of only astrophysical and instrumental backgrounds.

\section{\label{conclusion}Conclusion}

Our comprehensive analysis demonstrates Taiji's significant potential for detecting and characterizing SGWBs from cosmological FOPTs. Through systematic Bayesian analysis incorporating realistic instrumental noise and astrophysical foregrounds, we find that Taiji can robustly detect phase transition signals with energy densities exceeding $\Omega_{\mathrm{PT}} \gtrsim 1.4 \times 10^{-11}$ across most of its frequency band, with optimal sensitivity in the $10^{-3}$ to $10^{-2}$ Hz range. For stronger signals with $\Omega_{\mathrm{PT}} \gtrsim 1.1 \times 10^{-10}$, Taiji can determine the peak frequency with relative precision better than 10\%.
This sensitivity threshold represents a substantial improvement over current constraints~\cite{NANOGrav:2023gor,EPTA:2023fyk} and would enable tests of various early universe scenarios, including strongly supercooled transitions and those associated with composite Higgs models or hidden sector physics~\cite{Caprini:2019egz,Ellis:2020awk}. The consistency between our Bayesian evidence calculations and information-theoretic metrics provides a solid statistical foundation for future detection claims~\cite{Cornish:2015ikx}. While our study focused on the broken power-law spectral template, future work should explore more physically motivated spectral shapes directly connected to specific phase transition parameters such as transition temperature, strength, and bubble wall velocity~\cite{Hindmarsh:2020hop,Cutting:2020nla}. 

Our analysis employs the standard approach of combining multiple TDI channels (A, E, T) from a single detector. While this method effectively suppresses instrumental noise through the null channel T, it has inherent limitations for stochastic background detection~\cite{Muratore:2023gxh}. Single-detector analyses are fundamentally limited by the inability to distinguish between true GW signals and correlated instrumental artifacts. The null channel method, while useful for validation, cannot provide the same level of confidence as cross-correlation techniques between independent detectors.

For phase transition detection specifically, a multi-detector network could achieve detection thresholds potentially an order of magnitude lower than single-detector analyses, while providing more robust parameter estimation and reducing false positive rates. The different arm lengths and orientations of LISA (2.5 million km) and Taiji (3 million km) would offer complementary frequency responses, enhancing overall sensitivity across the millihertz band. The synergistic potential of Taiji operating concurrently with other space-based detectors like LISA would further enhance detection prospects through cross-correlation techniques~\cite{Orlando:2020oko,Liang:2021bde}.

\section*{Acknowledgments}
We thank the anonymous referee for providing constructive comments and suggestions that greatly improve the quality of this manuscript. We acknowledge the use of HPC Cluster of ITP-CAS.
ZCC is supported by the National Natural Science Foundation of China (Grant No.~12405056), the Natural Science Foundation of Hunan Province (Grant No.~2025JJ40006), and the Innovative Research Group of Hunan Province (Grant No.~2024JJ1006). 
QGH is supported by the grants from National Natural Science Foundation of China (Grant No.~12475065) and China Manned Space Program through its Space Application System.

\bibliography{ref}
\end{document}